\documentclass[superscriptaddress,author-year,onecolumn]{revtex4}
\usepackage{graphicx,subfigure,subeqnarray,fancyhdr,amsmath,epstopdf,multirow,color, mathrsfs,amssymb}
\usepackage[applemac]{inputenc}
\begin{document}

\renewcommand{\labelitemi}{$-$}
\newcommand{\change}[1]{{\color{black} #1}}
\newcommand{\Fc}{\mathcal{F}}\newcommand{\Rc}{\mathcal{R}}\newcommand{\dd}{\mathrm{d}}
\newcommand{\ee}{\mathrm{e}}\newcommand{\ci}{\mathrm{i}}\newcommand{\ib}{\mathbf{i}}
\newcommand{\jb}{\mathbf{j}}\newcommand{\kb}{\mathbf{k}}\newcommand{\ab}{\mathbf{a}}
\newcommand{\Fb}{\mathbf{F}}\newcommand{\fb}{\mathbf{f}}\newcommand{\Gb}{\mathbf{G}}
\newcommand{\Mb}{\mathbf{M}√Ñ}\newcommand{\nb}{\mathbf{n}}\newcommand{\Sb}{\mathbf{S}}
\newcommand{\Sbs}{\mathbf{S^*}}\newcommand{\Rb}{\mathbf{R}}\newcommand{\Sigb}{\boldsymbol{\Sigma}}
\newcommand{\Sigbs}{\boldsymbol{\Sigma^*}}\newcommand{\alphab}{\boldsymbol\alpha}
\newcommand{\omegab}{\boldsymbol{\omega}}
\newcommand{\sigmab}{\boldsymbol{\sigma}}
\newcommand{\epsb}{\boldsymbol{\epsilon}}
\newcommand{\ub}{\mathbf{u}}
\newcommand{\eb}{\mathbf{e}}\newcommand{\vv}[1]{\underline{#1}}\newcommand{\ev}{\vv{e}}
\newcommand{\rv}{\vv{r}}\newcommand{\TT}[1]{\underline{\underline{#1}}}\newcommand{\omb}{\mathbf{\omega}}
\newcommand{\Ub}{\mathbf{U}}\newcommand{\xb}{\mathbf{x}}\newcommand{\rb}{\mathbf{r}}
\newcommand{\ssb}{\mathbf{s}}\newcommand{\Xb}{\mathbf{X}}\newcommand{\Pe}{\mbox{Pe}}
\newcommand{\mean}[1]{\langle #1\rangle}
\newcommand{\ddp}{[p]^\pm}\newcommand{\taub}{\mbox{\boldmath$\tau$}}\newcommand{\Fr}{\mbox{\textit{Fr}}}
\let\grad\nabla\newcommand{\z}{\zeta}\newcommand{\kk}{\kappa}\newcommand{\tkk}{\tilde{\kappa}}
\newcommand{\e}{\varepsilon}\newcommand{\zb}{\bar{\zeta}}\let\grad\nabla\let\bcdot\cdot
\newcommand{\half}{{\textstyle\frac{1}{2}}}
\newcommand{\textfrac}[2]{{\textstyle\frac{#1}{#2}}}
\newcommand{\LF}[1]{{#1}^{\mathrm{LF}}}\newcommand{\Lap}[1]{{#1}^{\mathrm{L}}}
\newcommand{\ds}{*\!*}\newcommand{\cond}[2]{\frac{\mathrm{D} #1}{\mathrm{D} #2}}
\newcommand{\pard}[2]{\frac{\partial #1}{\partial #2}}\newcommand{\totd}[2]{\frac{\mathrm{d}#1}{\mathrm{d}#2}}
\newcommand{\pardd}[3]{\frac{\partial^2 #1}{\partial #2 \partial #3}}
\newcommand{\Rey}{\mbox{Re}}\newcommand{\Imag}{\mbox{Im}}
\newcommand{\Fpint}{=\!\!\!\!\!\!\!\int}
\newcommand{\txi}{\tilde\xi}\newcommand{\dxi}{\delta\xi}
\newcommand{\tpsi}{\tilde\psi}\newcommand{\dpsi}{\delta\psi}
\makeatletter
\def\sgn{\mathop{\operator@font sgn}}
\makeatother

\title{Fluid-solid-electric energy transport along piezoelectric flags}

\author{Yifan Xia}
\affiliation{LadHyX -- D\'epartement de M\'ecanique, Ecole Polytechnique -- CNRS, 91128 Palaiseau Cedex, France}
\author{Olivier Doar\'e}
\affiliation{IMSIA, ENSTA ParisTech, CNRS, CEA, EDF, Université Paris-Saclay, 828 bd des Maréchaux, 91762 Palaiseau Cedex, France} 
\date{\today}

\author{S\'ebastien Michelin}
\email{sebastien.michelin@ladhyx.polytechnique.fr}
\affiliation{LadHyX -- D\'epartement de M\'ecanique, Ecole Polytechnique -- CNRS, 91128 Palaiseau Cedex, France}
\begin{abstract}
The fluid-solid-electric dynamics of a flexible plate covered by interconnected piezoelectric patches in an axial steady flow are investigated using numerical simulations based on a reduced-order model of the fluid loading for slender structures. Beyond a critical flow velocity, the fluid-solid instability results in large amplitude flapping of the structure. Short piezoelectric patches positioned continuously along the plate convert its local deformation into electrical currents that are used within a single internal electrical network acting as an electric generator for the external output circuit. The relative role of the internal and external impedance on the energy harvesting of the system is presented and analyzed in the light of a full modeling of the electric and mechanical energy exchanges and transport along the structure. 

\end{abstract}
\maketitle

\section{Introduction}
Flow-induced vibrations have been extensively studied for the last 50 years: stemming from fundamental instabilities in the coupled dynamics of a moving solid body and a surrounding flow, they generate spontaneous, self-sustained and often large amplitude vibrations, that effectively convert some of an incoming flow's kinetic energy into solid kinetic or elastic energy~\citep{blevins1990,paidoussis2014}. Because of their critical and often damaging impact in industrial applications, most existing research has focused on the control of their linear dynamics in order to prevent the development of large amplitude vibrations~\citep{paidoussis1998,paidoussis2004}. The last decade has seen a renewed interest for these classical instabilities as energy harvesting systems, converting with an electric generator the vibration energy resulting from transverse galloping~\citep{barrero2010}, airfoil flutter~\citep{xiao2014}, vortex-induced vibrations~\citep{bernitsas2008} and axial flutter of flexible structures~\citep{singh2012b,michelin2013a}.

The latter, also known as ``flapping flag'' instability, is the result of the coupling of solid inertia and rigidity, to the destabilizing fluid forces resulting from the unsteady deflection of the flow by the moving structure~\citep{paidoussis2004,shelley2005}: beyond a critical flow velocity, large amplitude flapping develops, characterized by bending waves propagating along the plate~\citep{eloy2008,zhang2000,michelin2008}. Two main approaches have been proposed to harvest the associated energy: (i) the mechanical coupling of the flapping motion to a generator through its rotating mast~\citep{virot2016}, and (ii) the use of electro-active materials (e.g. piezoelectric materials) to directly convert the plate's deformation into an electric current~\citep{akcabay2012,giacomello2011,doare2011}. The present work focuses on the modeling of a piezoelectric flapping plate, for which an explicit description of the two-way electro-mechanical coupling and a more relevant definition of the harvesting efficiency have been obtained~\citep{doare2011,michelin2013a}, in contrast with empirical damping models for the harvesting process~\citep{tang2009,singh2012b}.

Modeling of such piezoelectric flags have so far followed two distinct routes: (i) a continuous approach, where the energy associated with the local bending is used  locally into independent circuits~\citep{doare2011,michelin2013a,xia2015a} and (ii) a discrete approach, where the structure is covered by a single element (or a small number) powering a single circuit~\citep{xia2015b,pineirua2015,xia2016,pineirua2016}. Beyond its formal simplicity, the main advantage of the former is its ability to exploit the entire structure's deformation, regardless of the deformation mode excited by the fluid-solid coupling. The latter is however the most  relevant for applications as it corresponds to a single output circuit, but the use of a single piezoelectric element effectively performs an average of the deformation, reducing the efficiency of the system~\citep{pineirua2015}.

The present work investigates an alternative approach that fully exploits the complex deformation of the structure, by using many short interconnected piezoelectric elements to create a single internal electrical network that can be connected to an external load. This electrical structure allows for the coupling of propagating bending and electrical waves, and richer electromechanical energy exchanges between the flapping flag and the output circuit.

The paper is organized as follows. In Section~\ref{sec:model}, the model and equations governing the dynamics of the piezoelectric flag are presented, in particular focusing on the original nonlocal circuit design and the resulting electromechanical exchanges within this fluid-solid-electric system. The resulting efficiency is then discussed in Section~\ref{sec:harvesting}, focusing in particular on the role of the circuit's properties. Building upon those results, Section~\ref{sec:fluxes} \change{analyzes} in detail the electrical energy fluxes along the flag, and potential routes of optimization of the harvester's design. Finally, conclusions and perspectives are presented in Section~\ref{sec:conclusions}.

\section{Fluid-solid-electric model of a piezoelectric flag}\label{sec:model}
\subsection{Description}
The energy harvester considered in this work is a thin inextensible flexible plate (or ``flag'') placed in an incoming uniform flow of velocity $U_\infty$ and density $\rho$, and covered by piezoelectric patches on each side. The plate is rectangular with dimensions $L$ and $H$ in the stream-wise and cross-flow directions, and its thickness is $h\ll H,\, L$. The piezoelectric plate assembly is supposed to have homogeneous structural properties and the effective mass per unit length and flexural rigidity are noted $\rho_s$ and $B$, respectively. The plate is clamped parallel to the flow at its leading edge, and is free to deform under the effect of its internal dynamics and of the flow forces. For simplicity, we condider here only purely planar deformations of the structure (i.e. twisting and cross-flow displacement are neglected).

The deformation of the plate periodically stretches and compresses the piezoelectric layers positioned on each side of the flag's surface, leading to a reorganization of their internal electrical structure and to an electric charge transfer between the electrodes of each patch. These patches are all identical and positioned by pairs (i.e. one patch on each side), shunted through the flag's surface; the polarities of the patches within each pair are reversed so that the effect of stretching of one patch and compression of the other during the flag's bending motion are additive~\citep{bisegna2006,doare2011}. The remaining two electrodes of each pair are connected to the electrical network (Figure~\ref{fig:circuit}). 

The electric state of the piezoelectric pair is characterized by the electric current and voltage between its free electrodes, noted respectively $\dot{Q}_i$ and $V_i$ for the $i$-th pair. The electro-mechanical coupling  is two-fold
\begin{itemize}
\item{\emph{a direct coupling:} the deformation of the flag induces a charge transfer so that
\begin{equation}
Q_i=CV_i+\chi[\theta(s_i^+)-\theta(s_i^-)],
\end{equation}
where $C$ is the internal capacitance of the patch pair, $s_i^\pm$ the Lagrangian coordinate of the leading and trailing edge of the patch along the flag's centerline, and $\chi$ the electro-mechanical coupling  that includes material and geometrical properties of the assemply \citep[see][]{bisegna2006,doare2011}}
\item{\emph{a reverse or feedback coupling:} the voltage within the pair induces an electric field inside the patch, resulting in a mechanical stress and an additional torque on the structure, $-\chi V_i$ applied between $s_i^-$ and $s_i^+$.}
\end{itemize}

\subsection{Piezoelectric coverage}
Our previous work on piezoelectric flags exclusively focused on local circuits: the energy extracted from the mechanical deformation is dissipated in an electric loop connected solely to that region, and there is no electrical energy exchange between different piezoelectric pairs. Such local circuits can take two forms: (i) one or a few patches cover the flag and energy is transferred to a small number of output circuits~\citep{xia2015b,pineirua2015,xia2016} or (ii) a large number of piezoelectric patches is considered so that a continuous limit can be used~\citep{doare2011,michelin2013a,xia2015a}. The advantage of the former is its simplicity and relevance to experiments (single output circuit). However, from a modeling point of view, this introduces discontinuities in the piezoelectric forcing on the flag; more importantly, the finite length of the piezoelectric patch effectively acts as an averaging filter in space: the forcing on the electric circuit is only a function of the change in orientation between $s_i^-$ and $s_i^+$, and not of the detailed bending. As a result, more energy can be harvested in the continuous limit consisting of many short piezoelectric patches and associated circuits, although a careful design of a finite number of a few piezoelectric patches allows to approach almost the same efficiency as that of the continuous limit~\citep{pineirua2015}.

\begin{figure}
\begin{center}
\includegraphics[width=12cm]{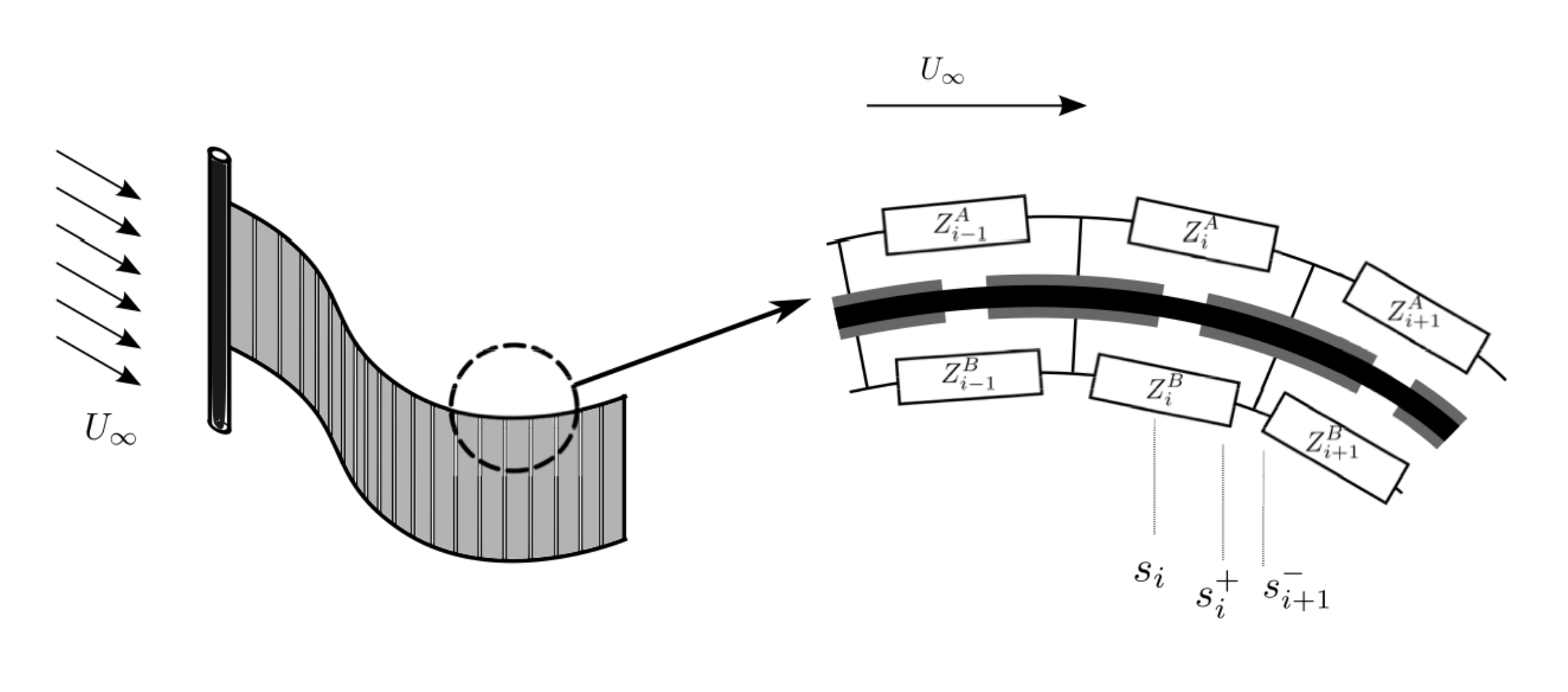}
\caption{Piezoelectric flag in a uniform flow. The surface of the flag is covered on both sides by piezoelectric patches (in grey) that are connected to their immediate neighbors.}\label{fig:flag}
\end{center}
\end{figure}

We consider here the alternative approach of interconnecting the different piezoelectric patch pairs electrically, so that energy can be transferred along the flag both mechanically and electrically. Adjacent pairs $i$ and $i+1$ are connected by two impedances (one on each side) $Z_i^A$ and $Z_i^B$ (Figure~\ref{fig:flag}). The advantage of this approach is twofold: (i) focusing on the limit of many small patches, providing a continuous coverage of the flag (i.e. $s_i^+=s_{i+1}^-$ and $s_i^+-s_i^-=\dd s\rightarrow 0$) allows for a maximum forcing of the circuit by removing any spatial average introduced by a finite patch length $l$; (ii) the integrated form of this connection provides the possibility to power a single output circuit with the entire apparatus by connecting the output load to the free electrodes located at the leading or trailing edge.

\begin{figure}
\begin{center}
\begin{tabular}{c}
\includegraphics[width=10cm]{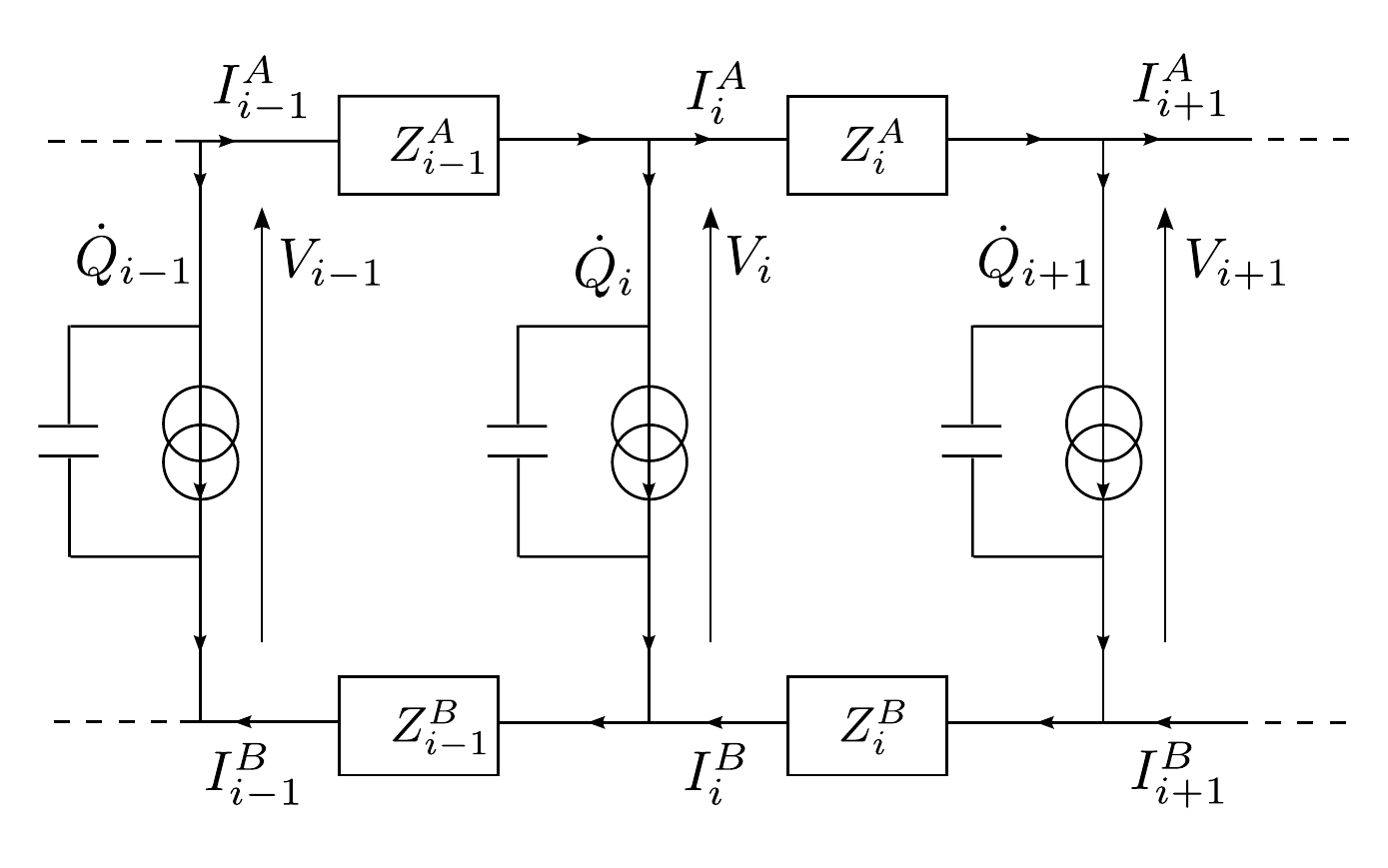}\vspace{.5cm}\\
\includegraphics[width=10cm]{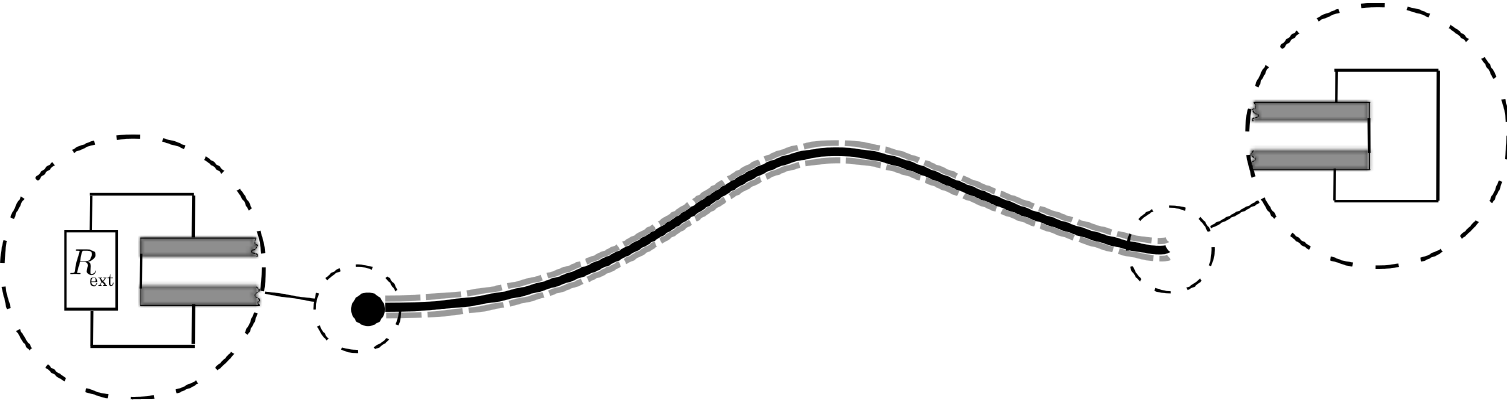}
\end{tabular}
\caption{(Top) Local electric circuit: each piezoelectric pair is equivalent from an electric point of view to a current generator and an internal capacitance. The current through a patch pair and the voltage at its free electrodes are respectively $V_i$ and $\dot{Q}_i$. (Bottom) Boundary conditions for leading edge harvesting, Eq.~\eqref{eq:elec_bcA}. }\label{fig:circuit}
\end{center}
\end{figure}

Applying Kirchhoff's circuit laws (Figure~\ref{fig:circuit}) leads to
\begin{align}
\dot{Q}_i=- I_i^A + I_{i-1}^A= - I_i^B + I_{i-1}^B, \label{eq:kirchhof1} \\
V_{i+1}-V_{i}=-Z_i^AI_i^A-Z_i^BI_i^B. \label{eq:kirchhof2}
\end{align}

\subsection{Continous model for the electrical network}
We follow here the approach presented in~\citep{doare2011}, by taking $\dd s\rightarrow 0$. We define $i_A$, $i_B$ and $v$ the continuous functions of $s$ such that $i_A(s_i)=I_i^A$, $i_B(s_i)=I_i^B$ and $v(s_i)=V_i$. Writing $q(s)$ the lineic charge transfer between the two layers of piezoelectric patches, $c$, $z_A$ and $z_B$ the lineic internal capacitance and internal impedance, the previous equations can be rewritten as
\begin{align}
q&=cv+\chi\pard{\theta}{s},\label{eq:continuous_direct}\\
\pard{^2 v}{s^2}&=z\cdot \dot{q}\label{eq:continuous_circuit},
\end{align}
with $z=z_A+z_B$. Equation~\eqref{eq:kirchhof1} indeed leads to $\pard{i_A}{s}=\pard{i_B}{s}=-\dot{q}$, or equivalently $i_A=i_B=i$ provided that the leading and trailing edges of the flag are not connected to each other by an outer circuit (i.e. there is not net current flowing through the flag). In that case, the  lineic impedance distribution between the two sides of the flag does not affect the dynamics and only their sum is relevant. In the following, we focus exclusively on a resistive connection between the different piezoelectric pairs, so that $z=r$ is the lineic resistance associated with the piezoelectric connection. The dynamics of the electrical circuit are therefore driven by
\begin{equation}
\pard{v}{t}-\frac{1}{rc}\pard{^2v}{s^2}+\frac{\chi}{c}\pard{^2\theta}{s\partial t}=0\label{eq:electric}
\end{equation}

\subsection{Equations of motion}
The additional piezoelectric torque applied on the structure now simply writes $-\chi v(s)$. An Euler--Bernoulli model is considered here to describe the two-dimensional motion of the piezoelectric plate
\begin{equation}
\rho_s\pard{^2\xb}{s^2}=\pard{}{s}\left[T\mathbf{t}-\nb\pard{}{s}\left(B\pard{\theta}{s}-\chi v\right)\right]+\mathbf{F}_\textrm{fluid}, \qquad \pard{\xb}{s}=\mathbf{t}.\label{eq:beam}
\end{equation}
In the previous equation, ($\mathbf{t}$,$\mathbf{n}$) are the local unit tangent and normal vectors in the plane of motion, $T$ is the tension within the structure and $\mathbf{F}_\textrm{fluid}$ is the fluid force on the plate. The second equation imposes the inextensibility of the flag. The flag is clamped at the leading edge and free at its trailing edge (the internal tension, torque and  shear force vanish). Therefore,
\begin{align}
 \textrm{at   } s=0,& \qquad \xb=0,\quad \theta=0,\label{eq:beam_bc1}\\
\textrm{at    }s=L,&\qquad T=B\pard{\theta}{s}-\chi v=B\pard{^2\theta}{s^2}-\chi\pard{v}{s}=0.\label{eq:beam_bc2}
\end{align}

\change{Up to this point, the fluid-solid-electric model is completely general, regardless of the method chosen to evaluate the fluid force on the flag $\mathbf{F}_\textrm{fluid}$.} Computing this fluid forcing can take many different routes, including direct numerical simulations of the viscous flow field~\citep{connell2007,banerjee2015},  and potential flow simulations using Panel Methods~\citep{qzhu2007}, point vortices~\citep{michelin2008} or vortex sheet models~\citep{alben2009}. In  the limit of a slender flag ($H\ll L$), an asymptotic model can be obtained for the inviscid local flow forces in terms of the local solid velocity using Lighthill's Large Amplitude Elongated Body Theory~\citep{lighthill1971}. This result based on the advection of fluid added momentum by the flow along the slender structures can also be interpreted (and proved) as an asymptotic expansion of the potential flow forces in the limit of small aspect ratio~\citep{candelier2011,candelier2013}. For freely-flapping bodies, this purely inviscid model must be complemented by a dissipative drag to account for the effect of lateral flow separation~\citep{candelier2011}. This physical feature of the flow field is described here by a quadratic drag associated with the normal displacement of the plate~\citep{taylor1952}. The result is a purely local formulation of the flow forces $\mathbf{F}_\textrm{fluid}$~\citep{singh2012b,eloy2012,michelin2013a},
\begin{equation}
\mathbf{F}_\textrm{fluid}=-\frac{\pi\rho H^2 m_a}{4}\left(\pard{(u_n\nb)}{t}-\pard{}{s}(u_t u_n\nb)+\frac{1}{2}\pard{(u_n^2\mathbf{t})}{s}\right)-\frac{1}{2}\rho c_d H |u_n|u_n\nb , \label{eq:fluid_model}
\end{equation} 
which is expressed solely in terms of the local relative velocity $\ub_r$ of the solid plate with respect to the background flow:
\begin{equation}
\ub_r=\pard{\xb}{t}-\mathbf{U}_\infty=u_t\mathbf{t}+u_n\nb.\label{eq:rel_vel}
\end{equation}
 In Eq.~\eqref{eq:fluid_model}, $m_a$ and $c_d$ are the added mass and drag coefficients. For the rectangular cross section considered here, $m_a=1$ and $c_d=1.8$. 

A main advantage of this method is that it doesn't require an explicit computation of the flow field which is embedded in Lightill's theory; this provides a strong reduction in the computational time, which is particularly convenient for large parametric or optimization analyses. This feature is also one of its main drawbacks, when dealing with multiple structures or confinement. A generalization of this method to deal with such configuration was recently proposed~\citep{mougel2016}.

\subsection{Output connection and energy efficiency}
The connectivity of adjacent piezoelectric pairs leaves two pairs of electrodes free at each end of the flag, that can be connected to an output circuit. In the following, we consider that the output circuit, namely a resistive load $R_\textrm{ext}$, is connected at one end of the flag, the other one being shunted (see Figure~\ref{fig:circuit}).
As a result, depending on the position of the harvesting circuit, the boundary conditions at the leading and trailing edges of the flag write: 
\begin{align}
\textrm{Leading edge harvesting:    }v(s=0)&=\frac{R_\textrm{ext}}{r}\pard{v}{s}(s=0)\quad\textrm{   and    }\quad v(s=L)=0,\label{eq:elec_bcA}\\
\textrm{Trailing edge harvesting:    }v(s=0)&=0\quad \textrm{   and   }\quad v(s=L)=-\frac{R_\textrm{ext}}{r}\pard{v}{s}(s=L).\label{eq:elec_bcB}
\end{align}

The output resistance is a proxy for the output circuit that uses the energy produced by the flag, therefore the output power of the system is defined as 
\begin{equation}
\mathcal{P}=\left\langle\frac{v_e^2}{R_\textrm{ext}}\right\rangle,
\end{equation}
 where $v_e$ is the voltage at the output resistance ($v_e=v(s=0)$ or $v(s=L)$ for a connection at the leading or trailing edge, respectively), and the efficiency $\eta$ of the system can be defined as 
 \begin{equation}
 \eta=\frac{\mathcal{P}}{\mathcal{P}_\textrm{ref}},\qquad \textrm{with    }\mathcal{P}_\textrm{ref}=\frac{1}{2}\rho U_\infty^3 H\mathcal{A},
 \end{equation}
namely, the ratio of the output power $\mathcal{P}$ to the kinetic energy flux $\mathcal{P}_\textrm{ref}$ through the surface occupied by the flag (here $\mathcal{A}$ is the peak-to-peak flapping amplitude at the trailing edge).

\subsection{Energy transfers along the flag}\label{sec:energy}
The flapping of a piezoelectric flag induces energy transfers between three different systems: the flowing fluid, the moving structure and the output electrical circuit. The conservation of mechanical energy is obtained by projecting Eq.~\eqref{eq:beam} onto the flag's local velocity
\begin{equation}
\pard{E_k}{t}+\pard{E_{el}}{t}=-\pard{\mathcal{F}_m}{s}-\mathcal{T}+\mathcal{W}_f,
\end{equation}
where $E_k=\rho_s|\partial \xb/\partial t|^2/2$ and $E_{el}=B(\partial\theta/\partial s)^2/2$ are the local kinetic and elastic energy densities on the flag, and
\begin{align}
\mathcal{F}_m&=-\pard{\xb}{t}\cdot\left[T\mathbf{t}-\pard{}{s}\left(B\pard{\theta}{s}-\chi v\right)\mathbf{n}\right]-\pard{\theta}{t}\left(B\pard{\theta}{s}-\chi v\right),\\
\mathcal{T}&=-\chi v\pard{^2\theta}{t\partial s},\\
\mathcal{W}_f&=\pard{\xb}{t}\cdot\mathbf{F}_\textrm{fluid},
\end{align}
are respectively the mechanical energy flux along the flag (i.e. the rate of work of internal forces and torques, measured positively from leading to trailing edge), the local rate of energy transfer from the flag to the circuit (solid-to-electric energy transfer), and the rate of work of the fluid forces (fluid-to-solid energy transfer).
The local conservation of electrical energy within each piezoelectric pair is obtained by multiplying the time-derivative of Eq.~\eqref{eq:continuous_direct} by $v$ and writes
\begin{equation}
\pard{E_C}{t}=\mathcal{T}-\mathcal{P}_{el},
\end{equation}
with $E_C=cv^2/2$ the energy stored in the piezoelectric capacitance, and $\mathcal{P}_{el}=-v\dot{q}$ the rate of energy transfer from the piezoelectric pairs to the circuit. Finally, for the nonlocal circuits considered here, Eq.~\eqref{eq:continuous_circuit} leads to
\begin{equation}
\mathcal{P}_{el}=\mathcal{P}_i+\pard{\mathcal{F}_{el}}{s},
\end{equation}
with   the electrical energy flux along the flag $\mathcal{F}_{el}$ measured positively from leading to trailing edge, and  the rate of dissipation of electrical energy in the internal resistors $\mathcal{P}_i$, respectively defined as
\begin{equation}
\mathcal{F}_{el}=-\frac{v}{r}\pard{v}{s}\quad\textrm{and}\quad\mathcal{P}_i=\frac{1}{2r}\left(\pard{v}{s}\right)^2.
\end{equation}

The mechanical boundary conditions on the flag imposed a fixed trailing edge and a free trailing edge, so that displacement or mechanical load vanishes at either end, in both rotation and translation. Therefore, $\mathcal{F}_m(s=0,L)=0$ (no flux of mechanical energy out of the flag). The electric boundary conditions, Eqs.~\eqref{eq:elec_bcA} or \eqref{eq:elec_bcB}, lead to $\mathcal{P}=-\mathcal{F}_{el}(s=0)$ (leading edge harvesting) or $\mathcal{P}=\mathcal{F}_{el}(s=L)$ (trailing edge harvesting). The electrical energy flux vanishes at the shunted extremity of the flag ($v=0$). Note that it would be the same for an open circuit condition ($\partial v/\partial s=0$).

\subsection{Non-dimensional equations}
Equations~\eqref{eq:electric}, \eqref{eq:beam} and \eqref{eq:fluid_model} together with boundary conditions \eqref{eq:beam_bc1}--\eqref{eq:beam_bc2} and \eqref{eq:elec_bcA}--\eqref{eq:elec_bcB} form a closed set of equations for the flag's position $\xb$, the internal tension $T$ and the voltage across the piezoelectric layers $v$. These equations are made non-dimensional using $L$, $L/U_\infty$, $\rho HL^2$ and $U_\infty\sqrt{\rho_s}{c}$ as characteristic length, time, mass and voltage. The problem is then completely determined by six non-dimensional parameters, namely
\begin{align}
H^*=\frac{H}{L},\qquad M^*= \frac{\rho H L}{\rho_s},\qquad U^*=U_\infty L\sqrt{\frac{\rho_s}{B}},\label{eq:nondimparam1}\\
 \alpha=\frac{\chi}{\sqrt{B c}},\qquad \beta=rcU_\infty L, \qquad \beta_\textrm{ext}=R_\textrm{ext}cU_\infty.\label{eq:nondimparam2}
\end{align}
$H^*$ is the plate's aspect ratio, and $M^*$ denotes the fluid-to-solid mass ratio: for large $M^*$ added mass effects dominate the solid inertia. $U^*$, the reduced velocity, is a relative measure of the destabilizing effect of flow forces on the flag and of the stabilization by internal rigidity. $\alpha$ is the coupling coefficient and scales both the direct and reverse coupling between the electrodynamic and mechanical problems. $\beta$ is the non-dimensional internal resistance of the circuit, and $\beta_\textrm{ext}$ the external reduced load of the output circuit. 

\section{Nonlinear dynamics and Energy harvesting}\label{sec:harvesting}
\subsection{Methods}
The non-dimensional form of Eqs~\eqref{eq:electric},\eqref{eq:beam} and \eqref{eq:fluid_model} and boundary conditions Eqs.~\eqref{eq:beam_bc1}--\eqref{eq:beam_bc2} and \eqref{eq:elec_bcA}--\eqref{eq:elec_bcB} are marched in time numerically using a second-order semi-explicit scheme~\citep{alben2009,michelin2013a} in order to obtain the dynamical position of the flag $\xb(s,t)$ and of the internal voltage $v(s,t)$. At a given instant $\tilde{t}$, the equations are recast as a set of nonlinear equations $\mathbf{F}(\mathbf{X})=0$, where $\mathbf{X}$ is a vector containing the discretized version of $\xb$ and $v$ at $\tilde{t}$. Integrals and derivatives in space are computed using a Chebyshev collocation method. The non-linear system is solved at each time step iteratively using Broyden's method~\citep{broyden1965}.

Initially, the internal piezoelectric capacitance is uncharged ($v=0$) and the flag is slightly displaced from its equilibrium position. Beyond a critical flow velocity, this perturbation is exponentially amplified by the fluid-solid-electric interactions and spontaneous flapping develops~\citep{michelin2013a,xia2015a}. The system is marched in time until a permanent saturated regime is achieved, for which time-averages can be defined without any ambiguity.

The energy harvesting efficiency is a function of six non-dimensional parameters listed in Eqs.~\eqref{eq:nondimparam1}--\eqref{eq:nondimparam2}. Previous publications have focused on the role of the inertia ratio $M^*$, on the relative importance of flow velocity and bending rigidity measured in $U^*$, on the coupling coefficient $\alpha$ and on the aspect ratio $H^*$ \citep{eloy2007,eloy2008,michelin2013a}. The goal of the present publication is to investigate the role of the  circuit's structure on the energy harvesting performance, and more specifically the effect of nonlocal electric coupling; in the following, we therefore focus on the influence of the reduced resistances $\beta$ and $\beta_\textrm{ext}$ on the harvesting performance. All simulations are thus performed for $H^*=0.5$, $\alpha=0.3$ and $U^*=15$, a value that is sufficiently above the critical flow velocity in the absence of piezoelectric coupling to avoid any restabilization of the structure due to the fluid-solid-electric interactions.

\subsection{Tuning and harvesting efficiency}
\begin{figure}
\begin{center}
\includegraphics[width=12cm]{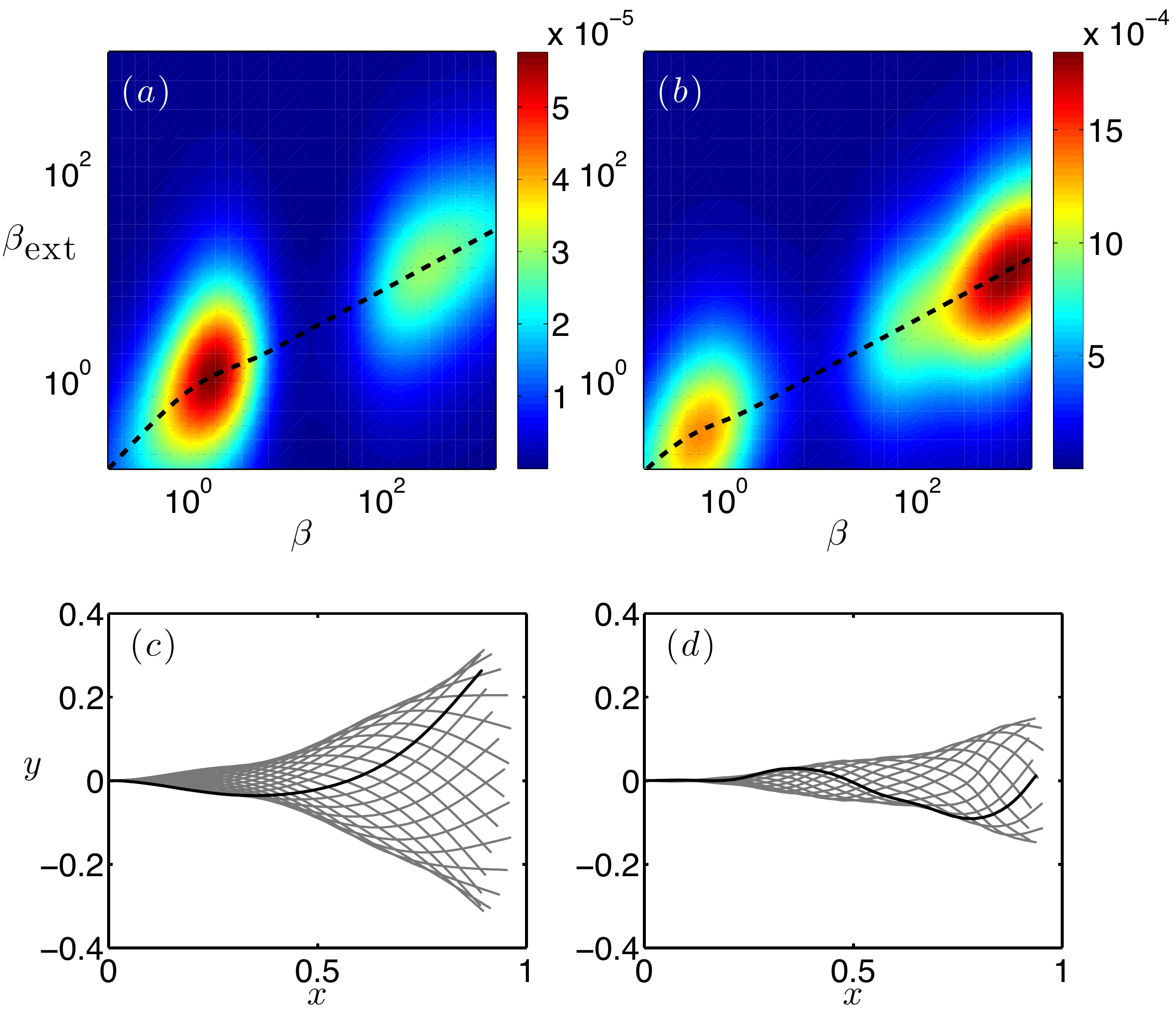}
\caption{(Top) Harvesting efficiency $\eta$ as a function of $\beta$ and $\beta_\textrm{ext}$ for (a) $M^*=1$ and (b) $M^*=10$, with a harvesting resistor positioned at the leading edge. The dashed line corresponds to the optimal impedance tuning condition, Eq.~\eqref{eq:optimal_tuning}. (Bottom) Flapping motion of the piezoelectric flag obtained for (c) $M^*=1$, $\beta=1.95$ and $\beta_\textrm{ext}=1.05$, and (d) $M^*=10$, $\beta=1.2\, 10^4$ and $\beta_\textrm{ext}=140$. The flapping frequency is measured as (c) $\omega=1.7U_\infty/L$ and $\omega=6.2U_\infty/L$ respectively. For all panels, $\alpha=0.3$, $H^*=0.5$ and $U^*=15$.}\label{fig:eta}
\end{center}
\end{figure}

Previous work on energy harvesting using piezoelectric flags has identified the critical role of the synchronization of the mechanical and electrical systems to maximize the energy transfers to the output resistance, whether for purely resistive circuits (tuning, \citep{michelin2013a}) or resonant circuits (\citep{xia2015a}). In the present case of nonlocal energy harvesting, Figure~\ref{fig:eta} identifies a non-trivial evolution of the efficiency with the internal and output resistances, and two optimal tuning regimes, namely for $\beta\sim\beta_\textrm{ext}=O(1)$ and for large but finite $\beta_\textrm{ext}$ and $\beta$. The position of these optimal configurations in the $(\beta,\beta_\textrm{ext})$-plane varies only weakly with the fluid-solid parameters (see in Figure~\ref{fig:eta} for the role of $M^*$ which plays a critical role in selecting the flapping mode shape), although the peak efficiency achieved in those configurations and their relative magnitude may change.

The non-dimensional parameters $\beta$ and $\beta_\textrm{ext}$ can be understood as ratios of an electric time-scale to the typical fluid-solid time scale associated with the fluid advection along the flag, and more generally the flapping frequency. When $\beta$ (resp. $\beta_\textrm{ext}$) is much lower or much greater than one, the internal (resp. output) resistance behaves as short or open circuit. 

The existence of an optimal configuration for finite $\beta$ and $\beta_\textrm{ext}$ is therefore expected. When $\beta_\textrm{ext}\ll 1$ or $\beta_\textrm{ext}\gg 1$, the output circuit effectively behaves as a short-circuit or open-circuit respectively, leading to either no voltage or current through the output circuit and no energy dissipation. Similarly, when $\beta\ll 1$, the internal resistor connecting neighboring piezoelectric patches effectively behave as short circuits, leading to a uniform voltage along the piezoelectric flag. The current powering the output resistance is proportional to $\partial v/\partial s$, therefore $\beta\ll 1$ results in negligible energy harvesting. Finally, when $\beta\gg 1$, the internal resistors effectively behave as open circuits, effectively disconnecting the different piezoelectric elements.  The output circuit is then only powered by the single closest patch, and for infinitesimal patches, leads to negligible efficiency.

\subsection{Tuning: a simplified model}
\label{sec:optimal_tuning}

The complexity of the problem comes here from the two-way coupling between the fluid, solid and electric dynamics. To rationalize the results presented above, we analyse a simpler problem, namely that of a prescribed flag kinematics. This is effectively equivalent to neglecting the effect on the flag's kinematics of the feedback coupling, or at least of the change in the feedback coupling introduced by varying the resistance parameters $\beta$ and $\beta_\textrm{ext}$; this is a good approximation in the limit of small $\alpha$.

\subsubsection{Optimal external tuning}
For simplicity, the flag's deformation is described as a traveling wave
\begin{equation}
\theta(s,t)=\Re\left[\Theta_0\ee^{\ci(ks-\omega t)}\right],
\end{equation}
with $\Re[\zeta]$ the real part of a complex number $\zeta$. The voltage in the circuit satisfies Eq.~\eqref{eq:continuous_circuit} together with boundary conditions, Eq~\eqref{eq:elec_bcA}. Writing $v(s,t)=\Re\left[f(s)\ee^{-\ci\omega t}\right]$, $f(s)$ is the unique solution of
\begin{equation}
f''+\ci\omega rc f=k\omega\chi r\Theta_0\ee^{\ci ks},\qquad f(0)=R_\textrm{ext}/r f'(0),\qquad f(L)=0.
\end{equation}
Writing $a=\sqrt{\ci \omega rc}=\sqrt{\omega rc/2}(1+\ci)$, $f(s)$ is obtained as
\begin{equation}
f(s)=\frac{k\omega\chi r\Theta_0}{a^2-k^2}\left[A\sin(as)+B\sin(a(s-L))+ \ee^{\ci ks}\right]
\end{equation}
with
\begin{align}
A&=-\frac{\ee^{\ci kL}}{\sin(aL)}\quad\textrm{and}\quad B=\frac{1+\gamma\left(\frac{aL\ee^{\ci kL}}{\sin(aL)}-\ci kL\right)}{\gamma aL \cos(aL)+\sin(aL)},
\end{align}
and $\gamma=\beta_\textrm{ext}/\beta$.
The total output power is then obtained as 
\begin{equation}
\mathcal{P}=\left\langle\frac{v(s=0)^2}{2R_\textrm{ext}}\right\rangle=\frac{|f(0)|^2}{2R_\textrm{ext}}=\frac{(k\omega\chi r\Theta_0)^2}{2R_\textrm{ext}(k^4+\omega^2 r^2c^2)}\left|1-B\sin(aL)\right|^2.
\end{equation}
After substitution, 
\begin{equation}
\mathcal{P}=\frac{r}{2\gamma L}\left(\frac{(k\omega\chi\Theta_0)^2 }{k^4+\omega^2r^2c^2}\right)\left|\frac{\gamma aL(\cos(aL)-\ee^{\ci kL})+\ci \gamma kL\sin(aL)}{\gamma aL \cos (aL)+\sin(aL)}\right|^2.
\end{equation}
Maximizing $\mathcal{P}$ with respect to the output resistance, all other dimensional quantities being held constant, is equivalent to maximizing $\gamma/|\gamma aL\cos(aL)+\sin(aL)|^2$ with respect to $\gamma$. It is easily shown that the optimal value for $\gamma$ is $\gamma_\textrm{opt}=|\tan(aL)/aL|$. Recalling that $aL=(1+\ci)\sqrt{\beta\bar\omega /2}$ (with $\bar\omega=\omega L/U_\infty$), this leads to an optimal relationship between $\beta_\textrm{ext}$ and $\beta$:
\begin{equation}
\left(\frac{\beta_\textrm{ext}}{\beta}\right)^2=\frac{1}{\beta\bar\omega}\left[\frac{\cosh(\sqrt{2\beta\bar\omega})-\cos(\sqrt{2\beta\bar\omega})}{\cosh(\sqrt{2\beta\bar\omega})+\cos(\sqrt{2\beta\bar\omega})}\right]=F(\beta\bar\omega).\label{eq:optimal_tuning}
\end{equation}
For each value of $\beta$, this optimal output tuning is shown on Figure~\ref{fig:eta} as a dashed line and coincides with the location of the two optimal configurations identified in the nonlinear simulations. Two regimes can be identified: (i) for $\beta\bar\omega\lesssim 1$, the optimal tuning of the internal and output impedance corresponds to $\beta_\textrm{ext}\sim\beta$ ($F\sim 1$), and the total internal resistance and output resistance are similar; (ii) for $\beta\bar\omega\gtrsim 1$, $\beta_\textrm{ext}\sim\sqrt{\beta/\bar\omega}$ and the internal resistance dominates ($F(\beta\bar\omega)\sim 1/\beta\bar\omega$).

This argument explains the existence of an optimal tuning between the output resistance ($\beta_\textrm{ext}$) and its internal counterpart ($\beta$),  and can be understood as an optimal matching of impedance between the continuous piezoelectric layer and the output connection. These results do not explain however why little energy is harvested for intermediate $\beta$ (regardless of $\beta_\textrm{ext}$).

\subsubsection{Avoiding internal dissipation}

\begin{figure}
\begin{center}
\includegraphics[width=12cm]{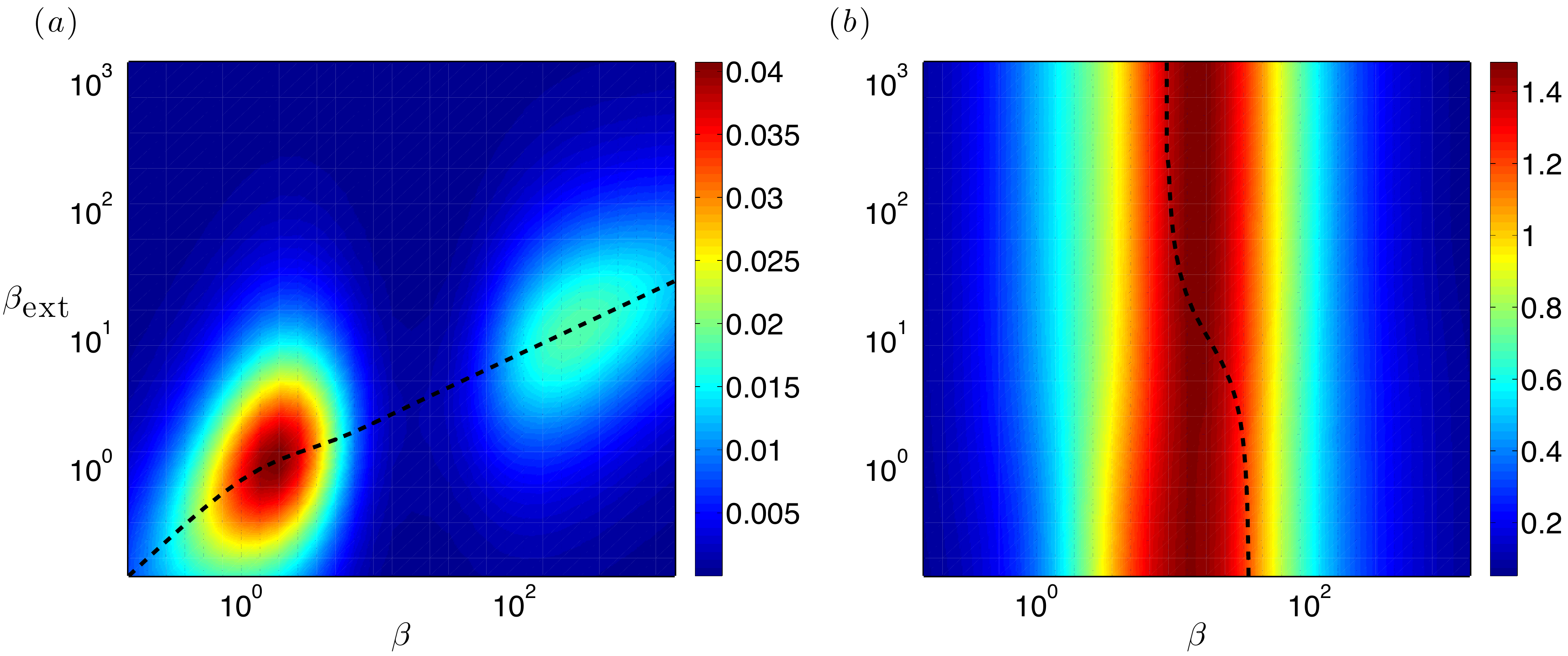}
\caption{Non-dimensional harvested power $\mathcal{P}$ \change{(a)} and internal dissipation $\mathcal{P}_\textrm{i}$ \change{(b)}t as a function of varying internal and external resistances $\beta$ and $\beta_\textrm{ext}$. For both panels, the dimensional powers are scaled by $\rho U_\infty^3HL$, and $M^*=1$, $\alpha=0.3$, $H^*=0.5$ and $U^*=15$ }\label{fig:Pint}
\end{center}
\end{figure}

To understand this second feature of Figure~\ref{fig:eta}, we turn back to the non-dimensional form of the electric equation, Eq.~\eqref{eq:continuous_circuit}. Its homogeneous part (i.e. without the piezoelectric forcing) reads
\begin{equation}
\pard{^2v}{s^2}-\beta\pard{v}{t}=0, \qquad v(0)=\frac{\beta_\textrm{ext}}{\beta}\pard{v}{s}(0),\qquad v(L)=0.
\end{equation}
which is formally equivalent to the heat equation. The characteristic time of the internal electrical network can be determined by searching for $v= \ee^{-t/\tau}V(s)$. After substitution in the equation above, this imposes that $\tau= \beta/\lambda^2$ with $\lambda$ solution of
\begin{equation}
\frac{\tan\lambda}{\lambda}+\frac{\beta_\textrm{ext}}{\beta}=0.
\end{equation}
Following \citep{michelin2013a}, we expect the dissipation to be maximum in the internal circuit when $\omega\tau\approx 2\pi$. When $\beta/\beta_\textrm{ext}\ll 1$ or $\beta/\beta_\textrm{ext}\gg 1$, $\lambda \approx \pi/2$ or $\pi$, respectively, which leads to $\omega\beta\sim \pi^3$. The frequency of flapping $\omega$ is essentially imposed by the flag motion, and this leads to a region of finite $\beta$ where dissipation in the internal circuit is maximum, leaving little energy available to the output circuit (Figure~\ref{fig:Pint}). Note that this $\beta$-range depends only weakly on $\beta_\textrm{ext}$.\\

The optimal harvesting conditions for nonlocal electric circuits can therefore be summarized as follows:
\begin{itemize}
\item{An optimal tuning of the internal and external impedances so that energy flowing to the harvesting end is entirely dissipated in the output resistor and only little energy is reflected.}
\item{A minimization of the internal dissipation by avoiding the perfect tuning condition between the flapping flag  and the internal circuit.}
\end{itemize}

\change{It should be noted that these conclusions are intrinsically linked to the general flapping pattern of the flag and more specifically the propagation of bending waves that act as a forcing mechanism on the circuit through the electro-mechanical coupling. The detailed fluid dynamics around the flag only plays a secondary role as exemplified by the agreement of the simulations and the results of simplified model. While a more complex representation of the flow field (e.g. using direct numerical simulations of the flow field) is likely to modify the exact details of the flapping pattern and the values of the harvested energy, the main results presented here, in particular the optimal harvesting conditions, would only be marginally modified.}

\section{Electric energy transfers along the flag}\label{sec:fluxes}
The previous results emphasize the critical role of energy transport along the nonlocal electrical circuit. In the analysis of energy transfers proposed in Section~\ref{sec:energy}, this corresponds to the electric flux $\mathcal{F}_{el}$ which is the rate of electrical energy transfer in the flow direction (left to right) at location $s$. Because the output resistor can not store electrical energy, the output power $\mathcal{P}$ is simply $-\mathcal{F}_{el}(0)$ (resp. $\mathcal{F}_{el}(L)$) for a resistance located upstream (resp. downstream).

\begin{figure}
\begin{center}
\includegraphics[width=9cm]{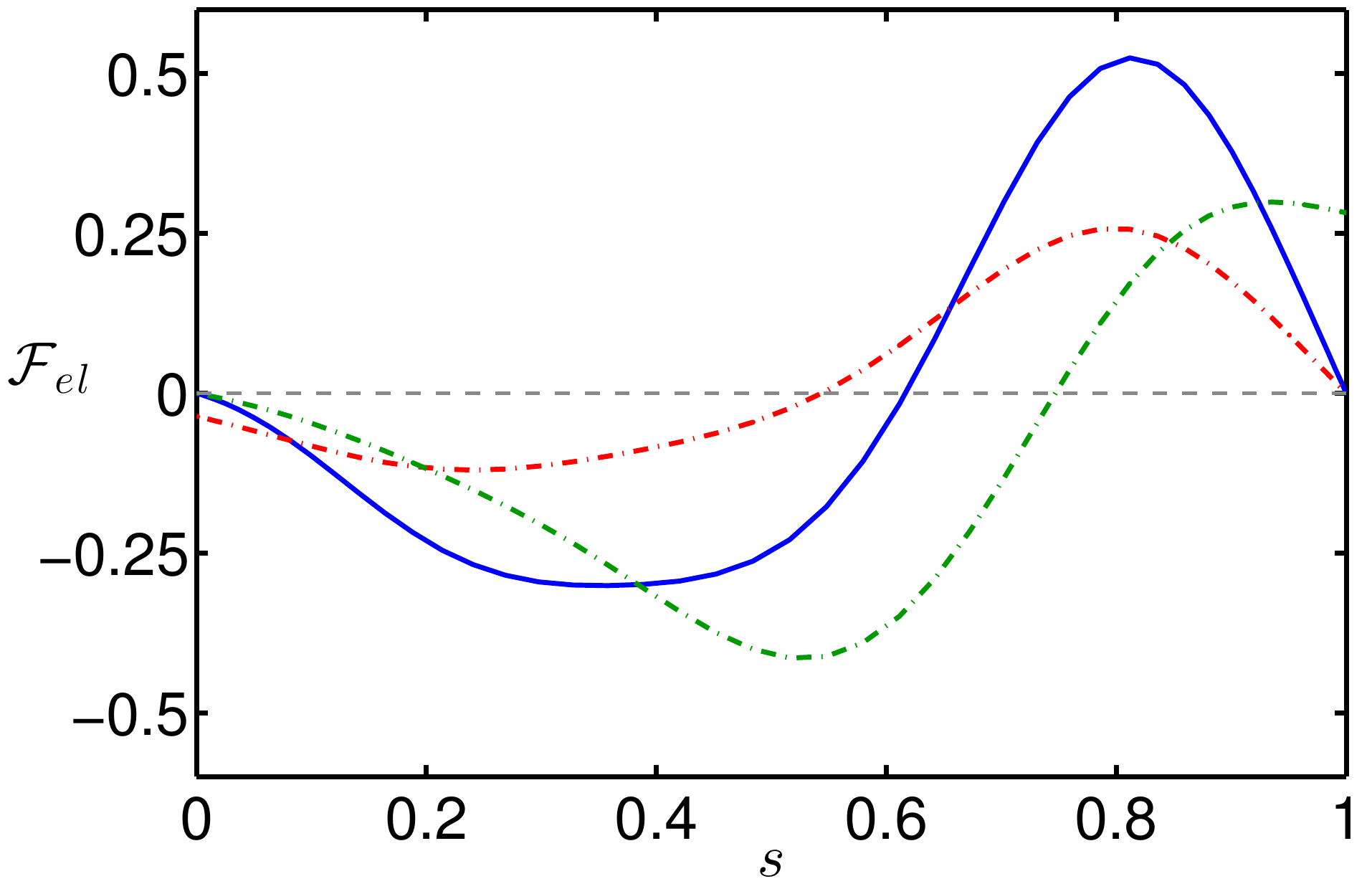}
\caption{Comparison of the electrical energy fluxes $\Fc_{\text{el}}$ with no harvesting resistor (blue) and for harvesting resistance at the leading or trailing ends (red and green, respectively) $\alpha=0.3$, $H^*=0.5$, $M^*=1$, $U^*=15$, $\beta=1.95$, $\beta_{\text{ext}}=1.05$ (optimal configuration on Figure~\ref{fig:eta}).\label{fig:fluxes}}
\end{center}
\end{figure}

In the absence of any output resistance, the electrical flux must vanish at both ends. Nevertheless, its variations indicate the amount of electrical energy transferred along the flag by the internal circuit (Figure~\ref{fig:fluxes}). One easily notes that the downstream half of the flag is characterized by an electrical energy transport in the direction of the flow and of the mechanical bending waves, while the upstream half is characterized by a reverse and lower energy transport \emph{against} the direction of the flow. At both ends of the flags, the electrical energy flux is therefore directed toward the flag's extremities. Since it must vanish there, energy must be either (i) returned to the mechanical system, and eventually the fluid flow, or (ii) dissipated in the output resistance.

The addition of an output resistance does not modify this general direction of transport of electrical energy $\mathcal{F}_{el}$, but significantly impacts its quantitative distribution, in particular in the vicinity of the harvesting extremity where $\mathcal{F}_{el}$ is not zero anymore, as shown on Figure~\ref{fig:fluxes}. The addition of an output resistance effectively relaxes the constraint $\mathcal{F}_{el}$ that imposed to dissipate or convert this energy when no output circuit was present: the energy can now be simply transferred to the output circuit.

This qualitative picture therefore suggests that an important insight on the optimal harvesting location can be gained from the distribution of electrical energy flux. Indeed, larger electrical energy flux at the boundary is equivalent to a larger output efficiency by definition, and Figure~\ref{fig:fluxes} suggests that one can determine the optimal location for the output circuit \emph{a priori} from the distribution of electrical energy flux in the absence of any harvesting: a greater amount of energy transport within the internal circuit  in the vicinity of one of the flag's extremity is likely to lead to greater efficiency once a harvesting resistance is added. This amounts to analyzing $\partial\mathcal{F}_{el}/\partial s$ near the boundary in the reference case.

\begin{figure}
\begin{center}
\begin{tabular}{cc}
\includegraphics[width=7cm]{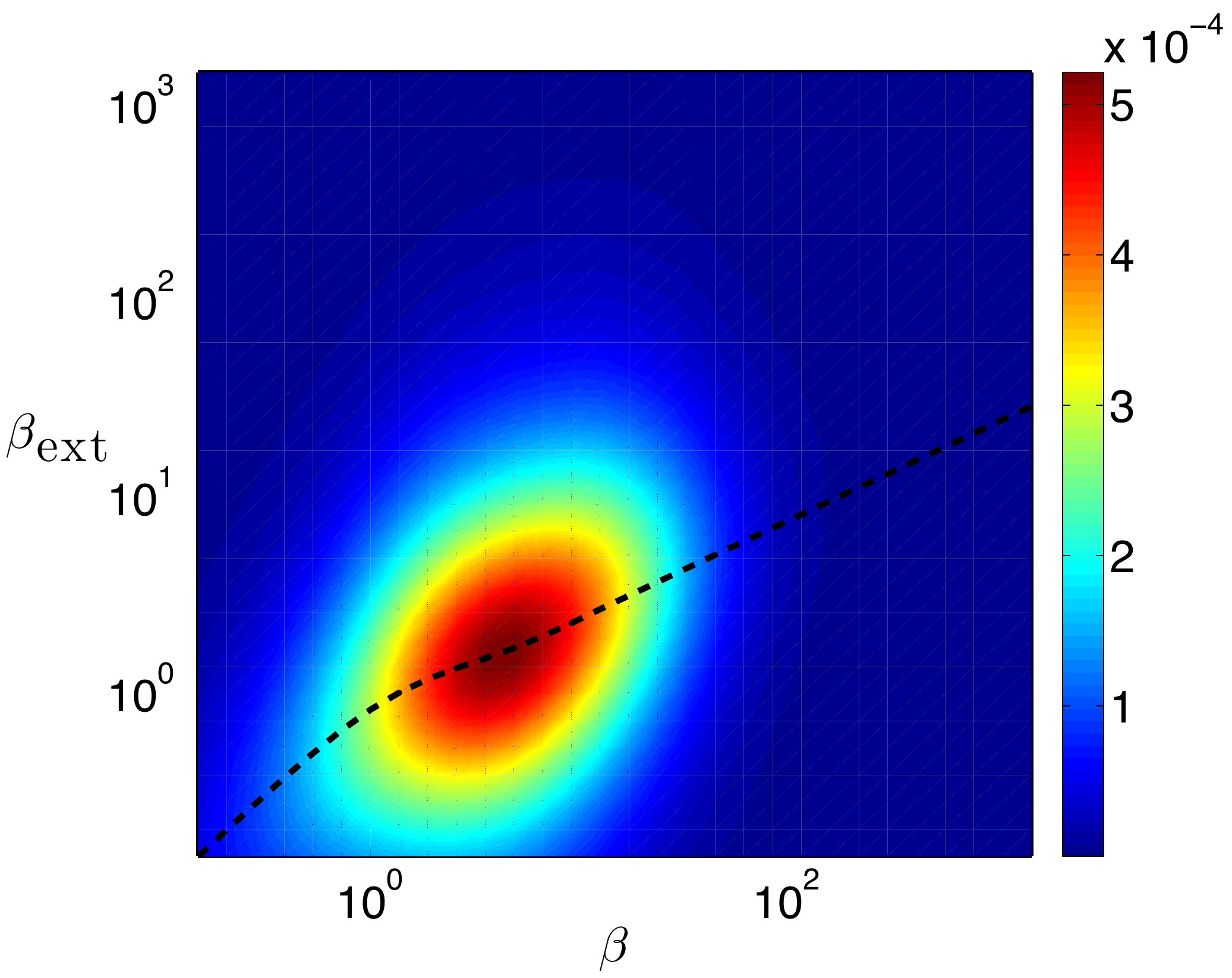} &
\includegraphics[width=6.7cm]{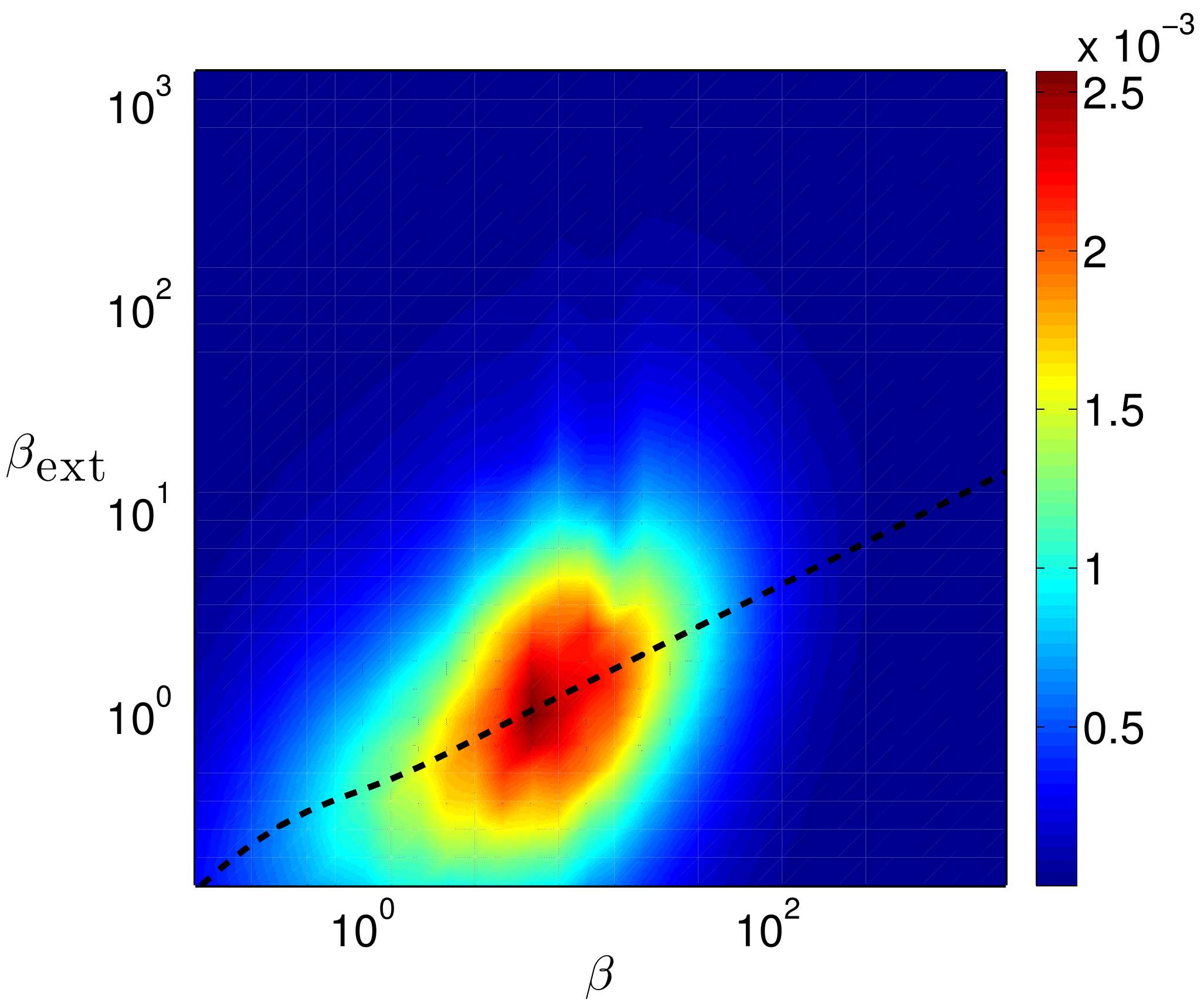} 
\end{tabular}
\caption{Harvesting efficiency $\eta$ as a function of $\beta$ and $\beta_\textrm{ext}$ for (a) $M^*=1$ and (b) $M^*=10$, with a harvesting resistor positioned at the trailing edge. The dashed line corresponds to the optimal impedance tuning condition identified in Eq.~\eqref{eq:optimal_tuning}. Here, $\alpha=0.3$, $H^*=0.5$ and $U^*=15$.}\label{fig:eta_trail}
\end{center}
\end{figure}

For the configuration considered in Figure~\ref{fig:fluxes} ($M^*=1$), this would suggest that trailing edge harvesting is more efficient, which is indeed confirmed by comparing the actual performance of both configurations  (Figures~\ref{fig:eta} and \ref{fig:eta_trail}). For $M^*=1$, the maximum efficiency obtained is an order of magnitude larger for trailing edge harvesting than what is obtained with a leading edge output circuit. Results obtained for larger $M^*$ (higher order flapping modes) show the same trend, but the gain is much less pronounced, suggesting a more complex mechanism. \change{For both $M^*$, a single peak is obtained in the harvested efficiency which lies on the theoretical prediction of the simplified tuning model, Eq.~\eqref{eq:optimal_tuning}. Repeating the analysis of section~\ref{sec:optimal_tuning} indeed shows that the optimal link between $\beta$ and $\beta_\textrm{ext}$ is not modified by moving the harvesting resistance to the trailing edge. The optimal value of $\beta$, and its relative position with respect to the region of maximum internal dissipation, is however modified, as well as the magnitude of the efficiency peak. The combination of these effects result in the existence of a single peak of efficiency (in contrast with two different peaks for leading-edge harvesting). }

\change{Furthermore, the distribution of electrical energy flux (Figure~\ref{fig:fluxes}) suggests} that alternative strategies may be even more efficient, namely by placing the harvesting resistance in the regions of maximum electrical energy flux. While beyond the scope of this study and modeling framework which focuses on a continuous model of the internal circuit, this  opens new opportunities in the optimal design of efficient harvesting systems.


\section{Conclusions}\label{sec:conclusions}
Powering an output external circuit from the flow-induced vibrations of a flexible structure requires dealing with a double complexity. On the mechanical side, flexibility allows for a continuous deformation and the solid's dynamics are characterized by a large number of degrees of freedom. Efficient energy harvesting requires to carefully analyze the effect of the extraction of energy on the flapping dynamics and on the energy transfers along the structure, often requiring a global optimization approach. On the electrical side, the continuous deformation of the structure must be exploited to produce a single electrical forcing to power the useful load. The approach presented here proposes a novel solution to deal with both challenges, by coupling the continous mechanical system to a continuous electrical system and exploit the energy exchanges between mechanical and electrical waves along the flapping structures. 

A minimal model for an output circuit was analyzed here, namely a single output resistance connected to one end of the flag. Optimal harvesting conditions were determined in terms of the characteristic output and internal impedance. Maximum energy transfer to the output circuit and maximum efficiency were obtained upon satisfying two different conditions: (i) an impedance tuning of the internal and output circuits to avoid reflection of energy, and (ii) an operating regime outside the range leading to maximum internal dissipation.

The analysis of the electrical energy transfers along the flag shows that energy harvesting is maximum when the output resistance is positioned near the flag's extremity where large electrical transport are present; in the absence of an output resistance, this energy needs to be either returned to the flow or dissipated internally, but the addition of an output circuit releases this constraint, and the available energy can be dissipated optimally in the harvesting circuit. 

This analysis suggests potential optimization routes for the positioning of the harvested circuit along the flag. This question is in fact critical for flow energy harvesting, beyond this particular geometry as demonstrated by several recent studies on energy harvesting using Vortex-Induced Vibrations of cables~\citep{grouthier2014,antoine2016}, and should be investigated in future work for piezoelectric flags.
\begin{acknowledgements}
This work was supported by the French National Research Agency ANR (Grant ANR-2012-JS09-0017). 
\end{acknowledgements}


\end{document}